\documentclass[fleqn,twoside]{article}
\usepackage{espcrc2}

\usepackage{graphicx}
\usepackage[figuresright]{rotating}

\def\amu{a_\mu}
\def\amuh{a_\mu^{{\mathrm{had}}}}
\def\MZ{M_Z}

\def\dah{\Delta\alpha^{(5)}_{\rm had}}
\def\dahs{\Delta\alpha^{(5)}_{\rm had}(s)}
\def\dahz{\Delta\alpha^{(5)}_{\rm had}(\MZ^2)}
\def\dah0{\Delta\alpha^{(5)}_{\rm had}(-s_0)}

\newcommand{\bmark}{\noi ~$\bullet$~~}
\newcommand{\power}[1]{\times 10^{#1}}
\newcommand{\be}{\begin{equation}}
\newcommand{\ee}{\end{equation}}
\newcommand{\ba}{\begin{eqnarray}}
\newcommand{\ea}{\end{eqnarray}}
\newcommand{\bea}{\begin{eqnarray*}}
\newcommand{\eea}{\end{eqnarray*}}
\newcommand{\bet}{\begin{center} \begin{tabular}}
\newcommand{\ent}{\end{tabular} \end{center}}
\newcommand{\sha}{\sigma(e^+e^- \rightarrow {\rm hadrons})}
\newcommand{\mz}{M^2_Z}

\newcommand{\al}{\alpha}

\newcommand{\dal}{\Delta \alpha}

\newcommand{\bary}{\begin{array}}
\newcommand{\eary}{\end{array}}
\newcommand{\noi}{\noindent}
\newcommand{\gapprox}{\raisebox{-.2ex}{$\stackrel{\textstyle>}
{\raisebox{-.6ex}[0ex][0ex]{$\sim$}}$}}

\newcommand{\crn}{\nonumber \\}
\newcommand{\nn}{\nonumber}
\newcommand{\gv}{\mbox{GeV}}

\newcommand{\bit}{\begin{itemize}}
\newcommand{\eit}{\end{itemize}}

\newcommand{\dalh}{\Delta \alpha^{\rm had}}
\newcommand{\epm}{e^+e^-}
\newcommand{\ra}{\rightarrow}
\newcommand{\sigh}{$\sigma(\epm \ra \mathrm{hadrons})$ }

\newcommand{\MSb}{$\overline{\rm MS}$ }
\newcommand{\alsMSb}{\alpha_s^{\overline{\rm MS}}}
\newcommand{\ha}{\frac12}
\newcommand{\sinf}{\sin^2 \Theta_f}

\newcommand{\sinW}{\sin^2 \Theta_W}
\newcommand{\cosf}{\cos^2 \Theta_f}

\newcommand{\cosW}{\cos^2 \Theta_W}
\newcommand{\cs}{ \;,\;\;}

\newcommand{\epo}{ \;.}

\newcommand{\aleffE}{$\alpha_\mathrm{eff}(E)$ }
\newcommand{\aleffZ}{$\alpha_\mathrm{eff}(M_Z)$ }

\newcommand{\tc}[1]{\multicolumn{1}{c}{#1}}

\newcommand{\AmS}{{\protect\the\textfont2
  A\kern-.1667em\lower.5ex\hbox{M}\kern-.125emS}}
\newfont{\liste}{pzdr scaled 1100}
\newfont{\grfett}{cmmib10 scaled 1100}

\title{Precision measurements of $\sigma_\mathrm{hadronic}$
for $\alpha_\mathrm{eff}(E)$ at ILC energies and $(g-2)_\mu$}

\author{F.~Jegerlehner\address{
 Humboldt-Universit\"at zu Berlin, Institut f\"ur Physik,
       Newtonstrasse 15, D-12489 Berlin, Germany}\address{
     Deutsches Elektronen-Synchrotron DESY,
       Platanenallee 6, D-15738 Zeuthen, Germany}
}
       
\begin{document}

\onecolumn{
\renewcommand{\thefootnote}{\fnsymbol{footnote}}
\setlength{\baselineskip}{0.52cm}
\thispagestyle{empty}
\begin{flushright} \begin{tabular}{c}
HU-EP-06/24\\ 
DESY 06-134 \\
SFB/CPP-06-38\\
August 2006 \end{tabular}
\end{flushright}

\setcounter{page}{0}

\mbox{}
\vspace*{\fill}
\begin{center}
{\Large\bf 
Precision measurements of $\sigma_\mathrm{hadronic}$} \\
\vspace{3mm}
{\Large\bf 
for $\alpha_\mathrm{eff}(E)$ at ILC energies and $(g-2)_\mu$}\\

\vspace{5em}
\large
F. Jegerlehner\footnote[1]{\noindent 
Work supported by DFG Sonderforschungsbereich Transregio 9-03
and in part by the European Community's Human Potential Program under
contract HPRN-CT-2002-00311 EURIDICE and the TARI Program under
contract RII3-CT-2004-506078.}
\\
\vspace{5em}
\normalsize
{\it Humboldt-Universit\"at zu Berlin, Institut f\"ur Physik,\\
       Newtonstrasse 15, D-12489 Berlin, Germany}\\

        and

{\it Deutsches Elektronen-Synchrotron DESY,\\
       Platanenallee 6, D-15738 Zeuthen, Germany}\\
\end{center}
\vspace*{\fill}}
\newpage

\begin{abstract}
A more precise determination of the effective fine structure constant
$\alpha_\mathrm{eff}(E)$ is mandatory for confronting data from future
precision experiments with precise SM predictions. Higher precision
would help a lot in monitoring new physics by increasing the
significance of any deviation from theory. At a future
$\epm$--collider like the ILC, as at LEP before,
$\alpha_\mathrm{eff}(E)$ plays the role the static zero momentum
$\alpha=\alpha_\mathrm{eff}(0)$ plays in low energy physics. However,
by going to the effective version of $\alpha$ one loses about a factor
$2 \times 10^{2}$ at $E=m_\mu$ to
$10^{5}$ at $E=M_Z$ in precision, such that for physics at the gauge
boson mass scale and beyond
$\alpha_\mathrm{eff}(E)$ is the least known basic parameter,
about a factor 20 less precise than the neutral gauge boson mass $M_Z$
and by about a factor 60 less precise than the Fermi constant
$G_F$. Examples of precision limitations are
$\alpha_\mathrm{eff}(m_\mu)$ which limits the theoretical precision of
the muon anomalous magnetic moment $a_\mu$ and
$\alpha_\mathrm{eff}(M_Z)$ which limits the accuracy of the prediction of the weak
mixing parameter $\sinf$ and indirectly the upper bound on the Higgs
mass $m_H$. An optimal exploitation of a future linear collider for
precision physics requires an improvement of the precision of
$\alpha_\mathrm{eff}(E)$ by something like a factor ten. We discuss a
strategy which should be able to reach this goal by appropriate
efforts in performing dedicated measurements of
$\sigma_\mathrm{hadronic}$ in a wide energy range as well as efforts
in theory and in particular improving the precision of the QCD
parameters $\alpha_s$, $m_c$ and $m_b$ by lattice QCD and/or more
precise determinations of them by experiments and perturbative QCD
efforts. Projects at VEPP-2000 (Novosibirsk) and DANAE/KLOE-2
(Frascati) are particularly important for improving on
$\alpha_\mathrm{eff}(M_Z)$ as well as $\alpha_\mathrm{eff}(m_\mu)$. Using
the Adler function as a monitor, one observes that we may obtain the
hadronic shift $\Delta \al _{\rm had}^{(5)}(\mz)$ as a sum $\Delta \al
_{\rm had}^{(5)}(-s_0)^{\mathrm{data}}$ + $\Delta \al _{\rm
had}^{(5)}(s_0,\mz)^{\mathrm{pQCD}}$ where the first term includes the
full non-perturbative part with the choice $s_0=(2.5~\gv)^2$ or larger. In such
a determination low-energy machines play a particularly important role
in the improvement program.  We present an up-to-date analysis
including the recent data from KLOE, SND, CMD-2 and BABAR. The
analysis based on $\epm$--data yields $\Delta \al _{\rm
had}^{(5)}(\mz) = 0.027593 \pm 0.000169$ [$\,\alpha^{-1}(\mz) = 128.938
\pm 0.023$] (splitting with $s_0=(10~\gv)^2$ to reduce
dependence on $m_c$), $\Delta \al _{\rm
had}^{(5)}(\mz) = 0.027607
\pm 0.000225$ [$\,\alpha^{-1}(\mz) = 128.947 \pm 0.035$] (standard approach),
and $\amuh=(692.1
\pm 5.6)\:\power{-10}$. The continuation of $\alpha_\mathrm{eff}(E)$
from the $Z$ mass scale to ILC energies may be obtained by means of
perturbative QCD. We emphasize the very high improvement
potential of the VEPP-2000 and DANAE/KLOE-2 projects. 

\vspace{1pc}
\end{abstract}

\maketitle

\section{INTRODUCTION}
\begin{figure}[h]
\vspace*{-3mm}
\centering
\includegraphics[height=4cm]{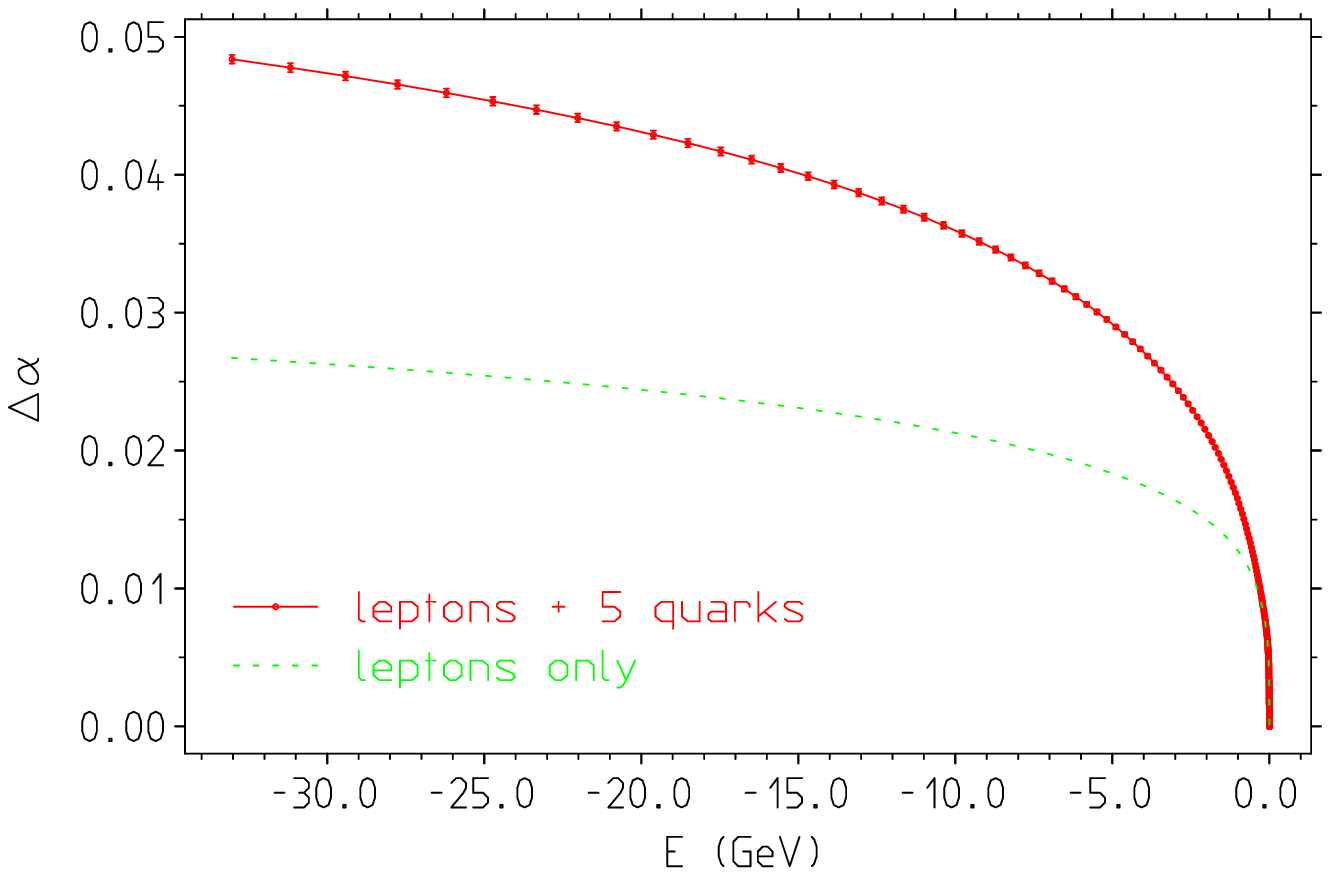}
\includegraphics[height=4cm]{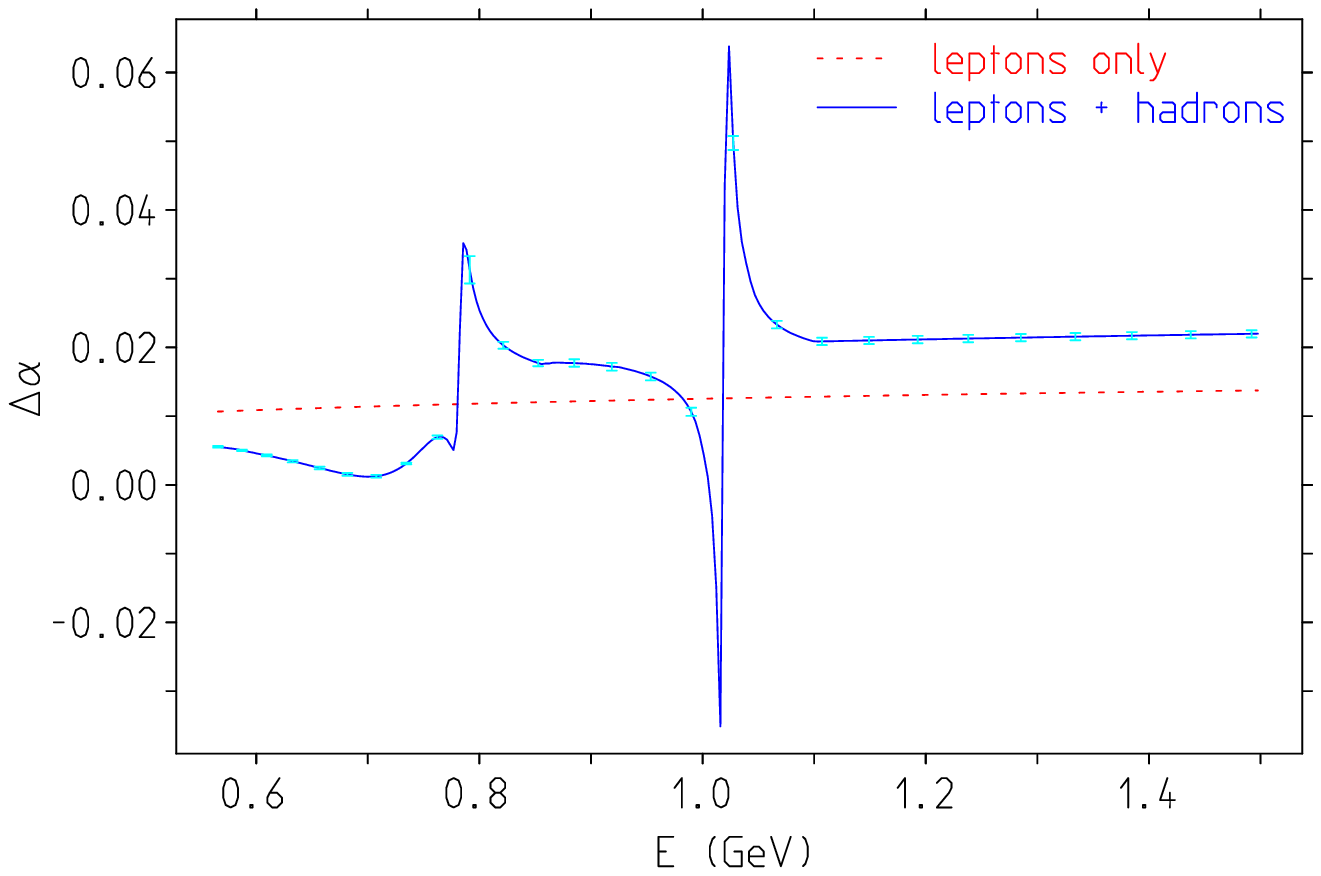}
\vspace*{-8mm}
\caption{The running of $\alpha(E)=\alpha(0)/(1-\Delta \alpha)$ in
terms of $\Delta \alpha$. The ``negative'' $E$ axis is
chosen to indicate space-like momentum transfer. The vertical bars at
selected points indicate the uncertainty. In the time-like region the
resonances lead to pronounced variations of the effective charge
(shown in the $\rho-\omega$ and $\phi$ region).}
\label{fig:runningalpha} 
\vspace*{-8mm}
\end{figure}

Of all non-perturbative hadronic effects entering
the predictions of a host of electroweak precision observables, the main
effect enters via the effective fine-structure ``constant''
$\alpha(E)$~\cite{FJLCnote01,TESLA01}. It is the non-perturbative
hadronic contribution to charge screening by vacuum polarization (VP)
which limits its precision. The value of $\alpha(E)$ is of interest for
a wide range of energies, well known examples are $\alpha(M_Z)$ or
$\alpha(m_\mu)$ where the latter accounts for the leading hadronic
contribution in the muon anomaly $a_\mu \equiv
(g-2)_\mu/2$ (see~\cite{EJ95,ADH98,DEHZ03,HMNT04,GJ04,ELZ03,TY04,BP05} for some recent
$\epm$--data based evaluations).  While electroweak effects like
lepton contributions are calculable
in perturbation theory, for the strong interaction part (hadron-
and/or quark- contributions) perturbation theory fails and a
dispersion integral over $\epm$--data encoded in \scriptsize
\bea
R_\gamma(s) \equiv \frac{\sigma(e^+e^- \rightarrow \gamma^*
\rightarrow {\rm hadrons})}{ \sigma(e^+e^- \rightarrow \gamma^* \rightarrow
\mu^+ \mu^-)}
\eea
\normalsize
provides a reliable approach to estimate the non-perturbative effects.
\begin{figure}[htb]
\vspace*{-4mm}      
\centering
\includegraphics[height=2.5cm]{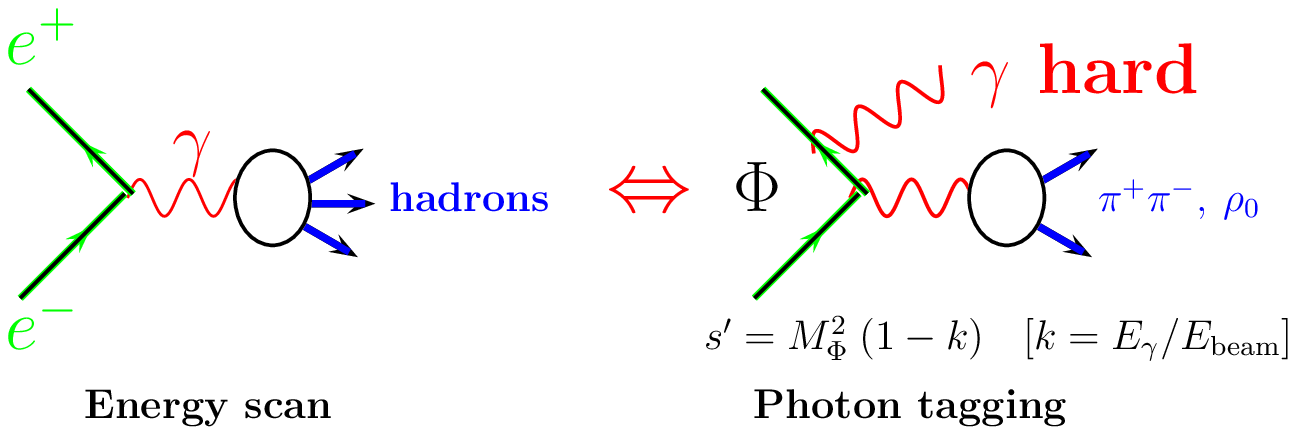}
\vspace*{-12mm}
\caption{The two schemes of measuring \sigh.}
\label{fig:SCvsRR} 
\vspace*{-5mm}
\end{figure}
\noi
Errors of data directly enter the theoretical uncertainties of any
prediction depending on $\alpha(E)$ at non-zero $E$. Evaluations of
the above-mentioned dispersion relation have developed into an art of
getting precise results still often from old cross--section
measurements of poor precision. What is needed is a reduction of the
present error by an order of magnitude within the next ten years. This
is a new challenge for precision experiments on \sigh such as ongoing
experiments KLOE, BABAR, and Belle which measure
$\sigma_\mathrm{hadronic}$ via radiative return or photon tagging (see
Fig.\ref{fig:SCvsRR})~\cite{RR,Benayoun:1999hm,Phokahra}.  For future
precision experiments, in many cases, we will need to know the running
$\alpha_\mathrm{em}$ very precisely, desirably at the per mill
level. Note that corrections are large and steeply increasing at low
$E$, as may be learned from Fig.\ref{fig:runningalpha}.

\begin{figure}[h]
\vspace*{-1mm}
\centering
\includegraphics[height=3cm]{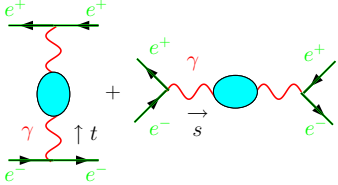}
\vspace*{-8mm}
\caption{Running $\alpha$ in Bhabha scattering as a normalization
process. The ``blobs'' are VP insertions into the tree level diagrams.}
\label{fig:bhabha} 
\end{figure}

An immediate question might be: why not measure $\alpha_\mathrm{eff}(E)$
directly, like the QCD running coupling $\alpha_s(s)$? The problem is that any
measurement requires a normalizing process like Bhabha
(Fig.\ref{fig:bhabha})
\noi
which itself depends on $\alpha_\mathrm{eff}(t)$ 
and $\alpha_\mathrm{eff}(s)$. In fact one is 
always measuring something like 
\ba
r(E) \propto \left(\alpha_\mathrm{eff}(s)/\alpha_\mathrm{eff}(t)\right)^2\cs
\ea
with $t=-\ha \: (s-4m_e^2)\:(1-\cos \theta)$. Unless one is
able to measure essentially at zero momentum transfer ($\sim 0$ angle),
because of the steep
rise of \aleffE at low energies a large fraction of the effect 
which we would like to determine drops out, especially the strongly rising
low-energy piece, which includes substantial non-perturbative effects
(see also~\cite{Trentadue:2006fx}).

\begin{figure}
\vspace*{-6mm}
\centering
\includegraphics[height=2cm]{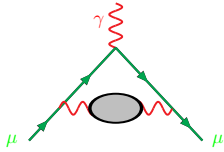}
\vspace*{-6mm}
\caption{Leading hadronic contribution to $g_\mu-2$}
\label{fig:gmudia} 
\vspace*{-3mm}
\end{figure}

It should be noted that not only \aleffE at high energies is of
interest. As a logarithmically increasing function the problems show
up at relatively low scales like for $\alpha_\mathrm{eff}(m_\mu)$,
which determines the leading uncertainty of the muon anomaly $\amu$
(see Fig.~\ref{fig:gmudia}). Another not so well known example is
$\alpha_\mathrm{eff}(M_\mathrm{proton})$ at the proton mass scale which
enters $\beta$--decay and affects the determination of the CKM element
$V_{ud}$~\cite{Sirlin94,Wilkinson:2003cp}.

In spite of the fact that the discussion presented here is not really
new, the substantial amount of new data in the low energy region from
KLOE, SND and CMD-2 and from BABAR leads to a noticeable change of the
error profile of \aleffE at different scales and it provides a strong
motivation for the upcoming VEPP-2000~\cite{VEPP2000} and future 
DANAE/KLOE-2~\cite{KLOE2} projects to continue precise measurements of
$\sigma_\mathrm{hadronic}$ in the region from threshold up to 2.5 GeV.

\section{$\alpha(M_Z)$ IN PRECISION PHYSICS}
At higher energies for all processes which are not dominated by a
single one photon exchange, $\alpha_\mathrm{eff}(E)$ enters in a
complicated way in ob\-ser\-vab\-les and cannot be measured in any direct
way. In the SM many parameters are interrelated as implied by the mass
generating Higgs mechanism. A nice example are parameters showing up
in typical four fermion and vector boson processes.  Unlike in QED and
QCD the SM is a spontaneously broken gauge theory which leads to a new
kind of parameter interdependencies, which have been tested at LEP with
high accuracy. Besides the fermion masses and mixing parameters the SM
has only 3 independent parameters the gauge couplings: $g$ and $g'$ and
the Higgs vacuum expectation value $v$, which may be fixed from the
most precisely measured quantities, namely, $\alpha$, $G_\mu$ and $M_Z$.
All other dependent parameters are then predictions which can be tested
and provide a moni\-tor for new physics.

The impact of the hadronic uncertainties of \aleffZ in physics of the
heavy gauge bosons $M_W$ and $M_Z$ are well known and for a more
detailed discussion I refer to my earlier articles~\cite{FJLCnote01}.
As mentioned before, in place of $\alpha$, which is known to a precision $\frac{\delta
\alpha}{\alpha} \sim 3.6 \times 10^{-9}$, the effective \aleffZ 
is needed and non-perturbative hadronic effects reduce its precision
to $\frac{\delta \alpha(M_Z)}{\alpha(M_Z)} \sim 1.6
\div 6.8 \times 10^{-4}$, while the other two basic parameters
are known with much better accuracy $\frac{\delta G_\mu}{G_\mu} \sim
8.6 \times 10^{-6}$ and $\frac{\delta M_Z}{M_Z} \sim 2.4 \times
10^{-5}$. The uncertainty in \aleffZ carries over to the $W$ mass
$M_W$ and to the weak mixing parameter $\sinf$ as
\scriptsize
\bea
\frac{\delta M_W}{M_W} &\sim& \ha \frac{\sinW}{\cosW-\sinW}
\;\delta \dal \sim 0.23 \;\delta \dal \\
\frac{\delta \sinf}{\sinf} &\sim& ~~\frac{\cosf}{\cosf-\sinf}
\;\delta \dal \sim 1.54 \;\delta \dal\;
\eea
\normalsize
and to all kinds of observables which depend on these parameters.  In
particular the indirect bounds on the Higgs mass obtained from
electroweak precision measurements are weakened thereby.

\section{EVALUATION OF $\alpha(M_Z)$}
\begin{figure}[t]
\vspace*{-4mm}
\centering
\includegraphics[height=4.2cm]{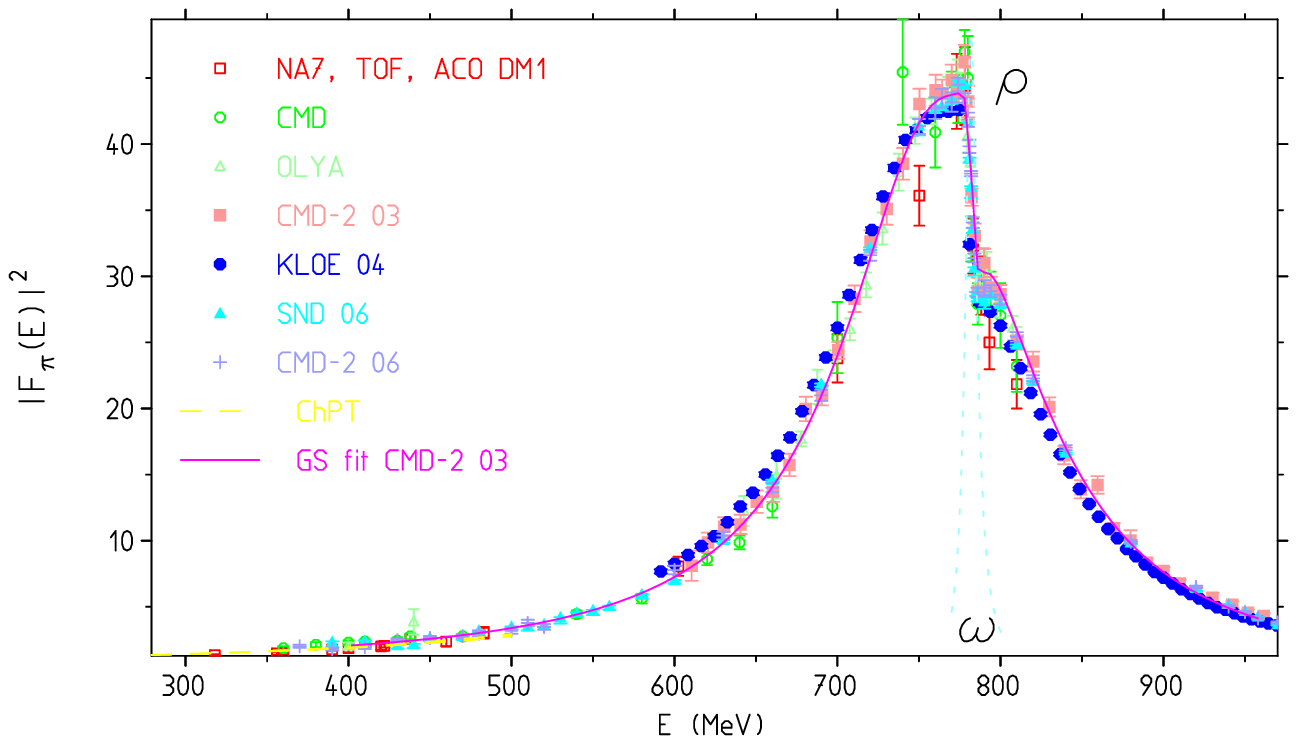}
\includegraphics[height=3.4cm]{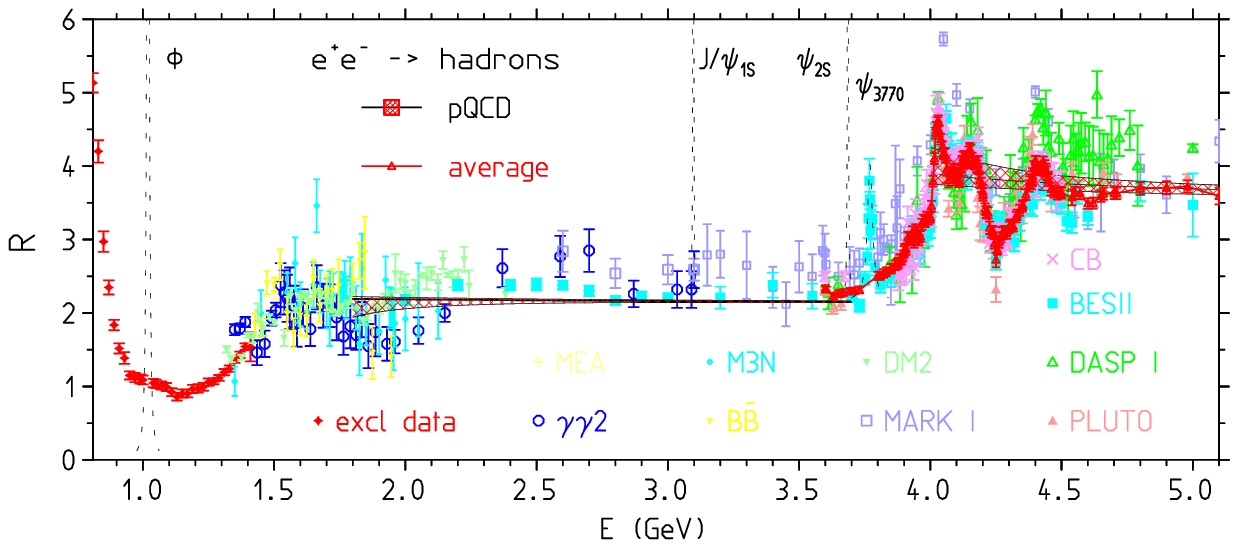}
\includegraphics[height=3.4cm]{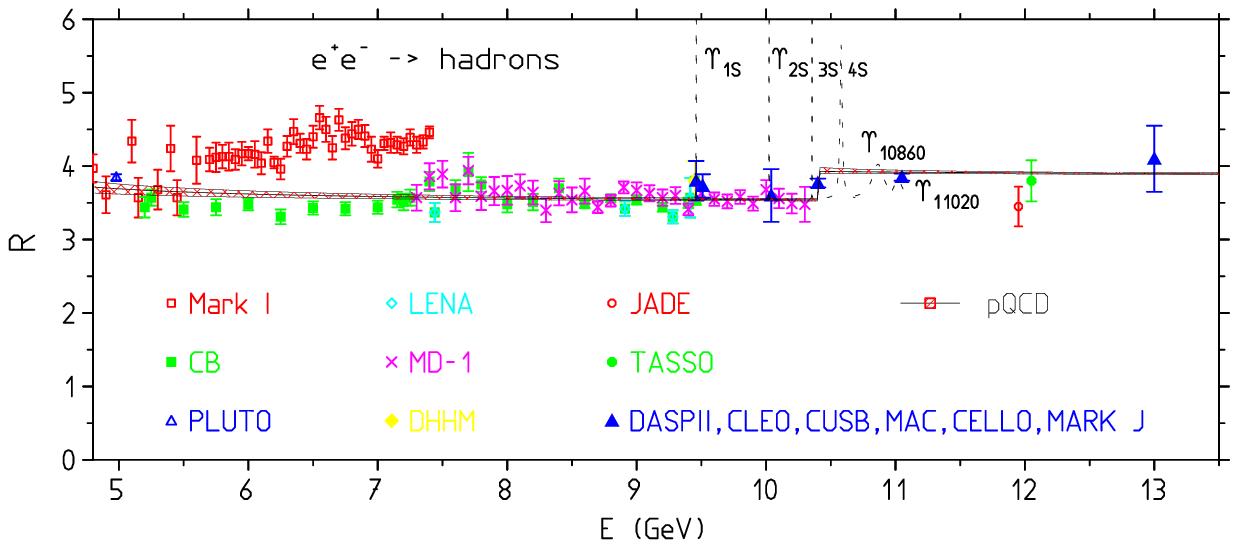}
\vspace*{-12mm}
\caption{A compilation of the presently available experimental
hadronic $\epm$--annihilation data}
\label{fig:Rdata} 
\vspace*{-7mm}
\end{figure}

 The non-perturbative hadronic shift $\dahs$, due to the 5 light quark
flavors, can be evaluated in
terms of $\sigma(e^+e^- \to {\rm hadrons})$ data via the well known
dispersion integral (see~\cite{EJ95} and references therein):
\ba
\dahs &=& - \frac{\alpha s}{3\pi}\;\bigg(\;\;\;
{\rm \footnotesize P}\!\!\!\!\!\!\!\!  \int\limits_{4m_\pi^2}^{E^2_{\rm cut}} ds'
\frac{{R^{\mathrm{data}}_\gamma(s')}}{s'(s'-s)}
\crn &&~~~~~~~~+ {\rm \footnotesize P}\!\!\!\!\!\!\!\!
\int\limits_{E^2_{\rm cut}}^\infty ds'
\frac{{ R^{\mathrm{pQCD}}_\gamma(s')}}{s'(s'-s)}\,\, 
\bigg)
\label{aleffcut}
\ea
where
\ba
R_\gamma(s) \equiv \frac{\sigma^{(0)}(e^+e^- \rightarrow \gamma^*
\rightarrow {\rm hadrons})}{\frac{4\pi \alpha^2}{3s}} \epo 
\label{Rundressed}
\ea
A compilation of the data is shown in Fig.~\ref{fig:Rdata}.  

\begin{figure}
\vspace*{-00mm}
\centering
\includegraphics[height=2.9cm]{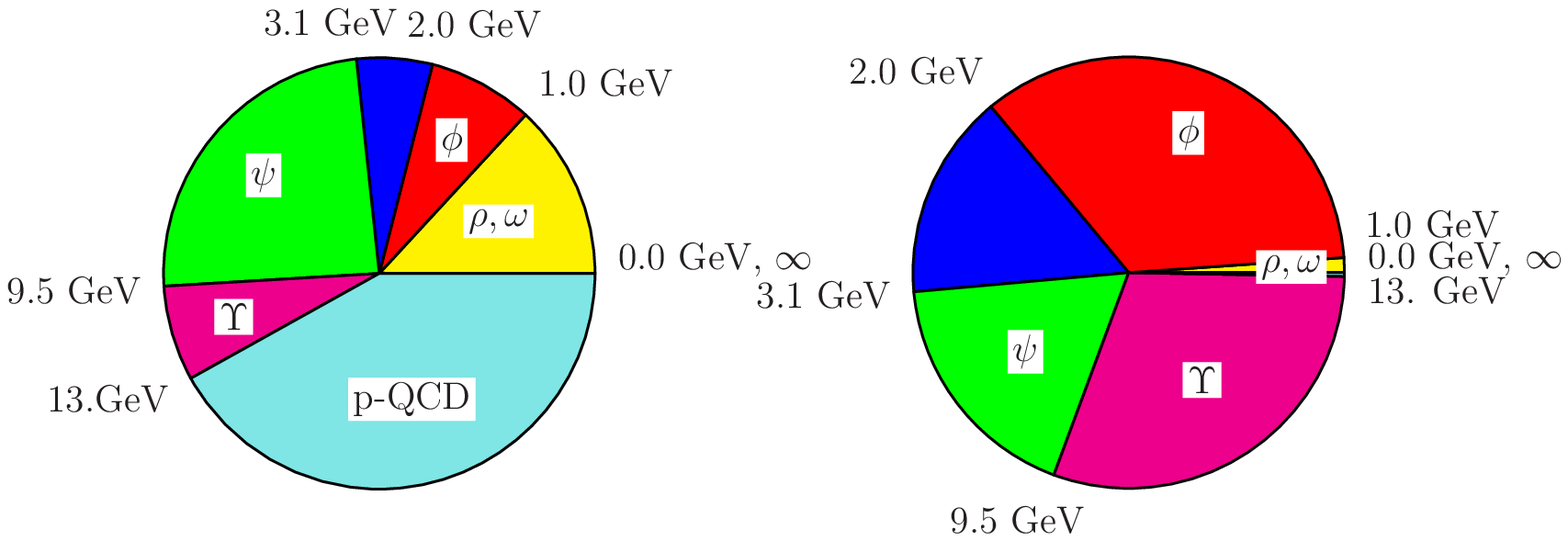}
\vspace*{-16mm}
\caption{$\dahz$: contributions (left) and errors$^2$ (right) from different regions}
\label{fig:distR} 
\vspace*{-4mm}
\end{figure}
\begin{figure}
\vspace*{-3mm}
\centering
\includegraphics[height=2.9cm]{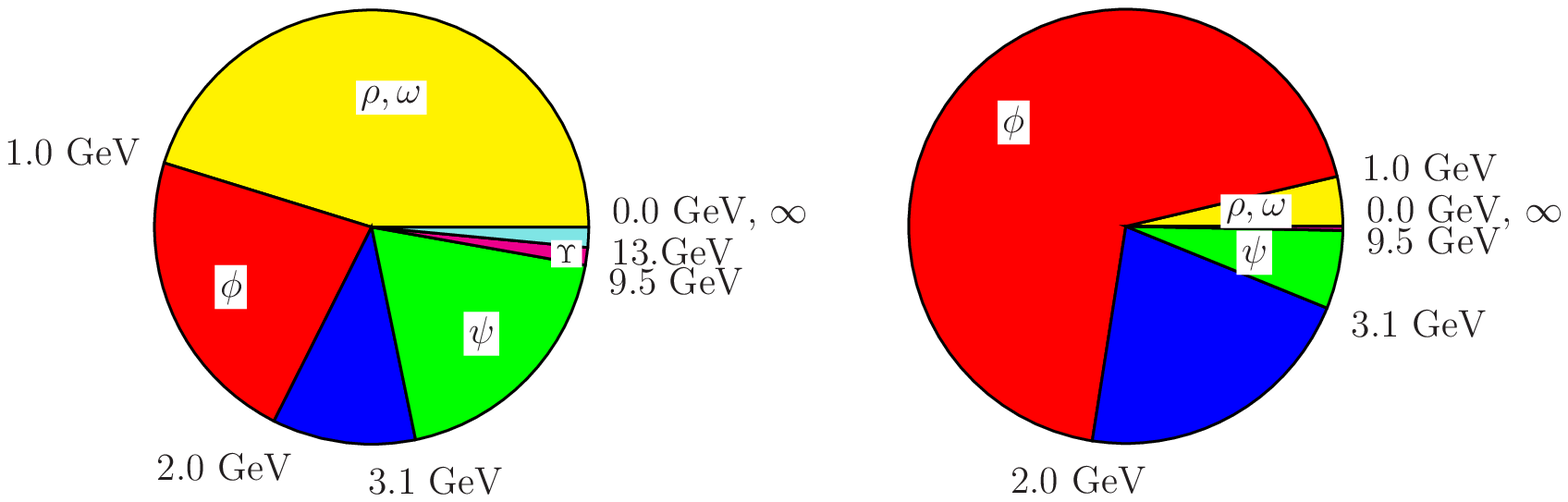}
\vspace*{-14mm}
\caption{$\dah0$: contributions (left) and errors$^2$ (right) from different regions}
\label{fig:distD} 
\vspace*{-4mm}
\end{figure}

\noi
For the evaluation at $M_Z=$ 91.1876 GeV we take

\bmark $R(s)$ data up to $\sqrt{s}=E_{cut}=5.2$ GeV and 
for $\Upsilon$ resonances region between 9.6 and 13 GeV,

\bmark perturbative QCD~\cite{GKL,ChK95} from 5.2 to 9.6 GeV
and for the high energy tail above 13 GeV, as recommended in~\cite{HS02}.
As a result we obtain
\ba
\Delta \al _{\rm hadrons}^{(5)}(\mz) & = & 0.027607 \pm 0.000225 \crn
\alpha^{-1}(\mz) & = & 128.947 \pm 0.035\:.   
\ea
Contributions from various energy regions and the origin of the errors
are shown in Fig.~\ref{fig:distR}.

As we will explain below $\dahz$, or more generally \aleffE for
$E>2.5~\gv$, may be evaluated in a different way, by exploiting pQCD
as much as possible in a well controlled fashion via the Adler function.
In this Adler function approach the complete non-perturbative part may
be evaluated at $\sqrt{s_0}=2.5~\gv$ where 
\bea
\Delta \al _{\rm had}^{(5)}(-s_0) & = & 0.007364 \pm 0.000101\:.
\eea
and $\dahz-\dah0$ is reliably calculable using pQCD.
The profile of $\dah0$ is shown in Fig.~\ref{fig:distD}; for more details
we refer to Table \ref{tab:alphacont} below.

\section{A LOOK AT THE $\epm$--DATA}
In order to learn where results have to be improved we briefly have a
closer look at the existing $\epm \to \mathrm{hadrons}$ cross-section
data. Since our analysis~\cite{EJ95} in 1995 data from
MD1~\cite{MD196}, BES-II~\cite{BES} and from CMD-2~\cite{CMD2} have
lead to a substantial reduction in the hadronic uncertainties on
$\dalh$ and $\amuh$. More
recently KLOE~\cite{Aloisio:2004bu}, SND~\cite{Achasov:2006vp} and
CMD-2~\cite{Aulchenko:2006na} published new measurements in the region
below 1.4 GeV. Unfortunately the agreement between the different
experiments is not very satisfactory. Nevertheless, the progress is
substantial, the low energy domain is no longer essentially
dominated by one experiment (CMD-2 2003) and the errors of the
combined data are reduced noticeably and existing problems can be
settled by ongoing experiments. In the past few years also more data
on the purely neutral channels $\pi^0 \gamma,\pi^0\pi^0 \gamma,~\eta
\gamma,~\pi^0\eta \gamma,~\eta' \gamma,~\omega\pi^0$ mainly from
SND~\cite{pigammaetcSND} and CMD-2~\cite{pigammaetcCMD} have been
included. Substantial improvement on the $\omega$ and $\phi$
resonances, on $4\pi$ and $5\pi$ channels and a study of $\eta \pi^+
\pi^-,~\omega \pi^+ \pi^-$~\cite{etapipiCMD} contributed to the
improvement. An important conclusion from the CMD-2
study~\cite{pigammaetcCMD} is that the $\omega$- and $\phi$-resonance decays completely
saturate the cross sections of these channels. Consequently, there are
no unaccounted contributions to the $R$ value from such
neutral channels (see~\cite{Eidelman:2005tk} for a more detailed
expert summary). Further data included $3\pi$, $4\pi$, $KK$ and $pp$
channels and come from SND, CMD-2, BES-II and 
CLEO~\cite{Achasov:2002ud,Achasov:2003bv,Akhmetshin:2004dy,Akhmetshin:1999ym,BES05,Pedlar:2005sj}

As in other more recent evaluations, modes not measured directly have
been included by estimating them using isospin relations and known
branching fractions of decay modes. Included are (see also~\cite{DEHZ03,HMNT04})
{\bf (1)}
$\pi^+\pi^-3\pi^0 \sim [2(\pi^+\pi^-)\:\pi^0 - \eta
\pi^+\pi^-)]/2 + \eta \pi^+\pi^-\,\cdot\mathrm{BR}(\eta \to 2\pi^0)$,
{\bf (2)} $\eta \pi \pi$ not included already in the $5\pi$ modes, 
{\bf (3)} $\omega
(\omega \to \pi^0 \gamma) \pi \pi$ including the $\omega \pi^0 \pi^0$
mode via isospin (= $(\omega \pi^+\pi^-$ + $\omega \pi^0\pi^0
[=1/2 \omega \pi^+\pi^-])
\,\cdot\mathrm{BR}(\omega \to \pi^0\gamma$),
{\bf (4)} unseen $KK\pi\pi$ modes using $\epm \to K^0_S X$ data of DM1, which
amounts to taking 2 times the total for the modes $K^0 \bar{K}^0
(\pi\pi)^0$ and $K^+ K^- (\pi\pi)^0$ (=2($K^0_S X - K^0_S\bar{K}_L^0 - K^0_S
\bar{K}^\pm\pi^\mp - K^+ K^- \pi^0)$), $K^+K^-\pi^+\pi^-$ measured by
DM1 and BABAR is evaluated from the data and hence is to be subtracted from
the total $KK\pi\pi$ contribution, {\bf (5)} missing isospin
modes in $6\pi$: $\pi^+ \pi^- 4\pi^0$ ($\simeq 0.093\cdot
3(\pi^+\pi^-)+0.031\cdot 2(\pi^+\pi^-)\:2\pi^0$),
{\bf (6)}
$P\gamma$ ($P=\pi^0,\eta$) (note $\pi^0\pi^0\gamma$ is a version of
$\omega \pi^0$ already included), {\bf (7)} $K^+ K^- \pi^0$ (was missing in
non-DM2 R compilations) also accounting for $K^0_L K^0_S \pi^0[= K^+
K^- \pi^0]$. Channel measured for the first time include $K^+K^-\,
2(\pi^+\pi^-)$ and $2(K^+K^-)$ (BABAR). For a useful review on the $\epm$ data 
we refer to~\cite{WhalleyR03} (up to 2003).

\begin{figure}[t]
\centering
\includegraphics[height=8cm]{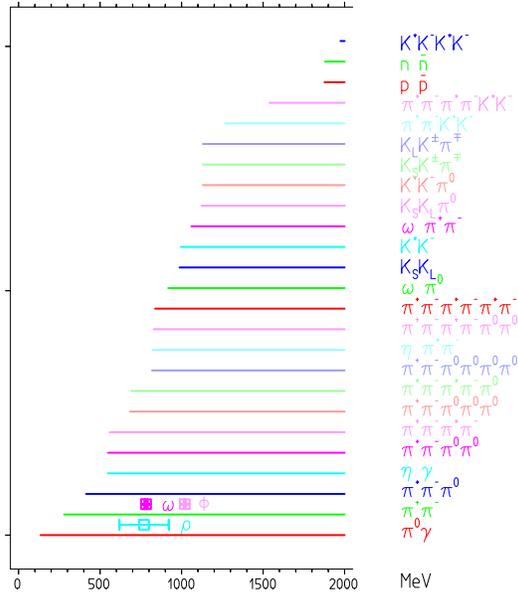}
\vspace*{-4mm}
\caption{Thresholds for exclusive multi particle channels below 2 GeV}
\vspace*{-4mm}
\label{fig:multi} 
\end{figure}

\begin{figure}[t]
\vspace*{-4mm}
\centering
\includegraphics[height=5cm]{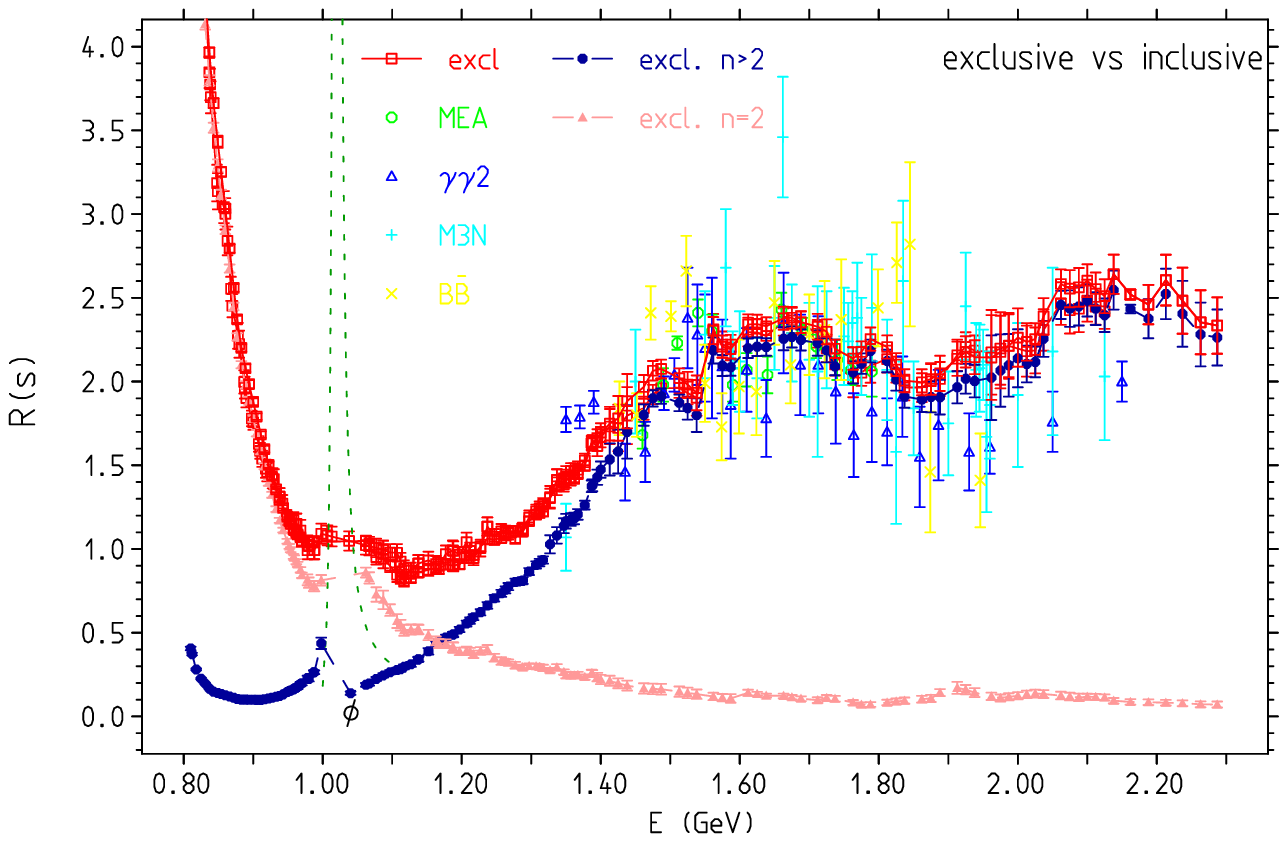}
\includegraphics[height=5cm]{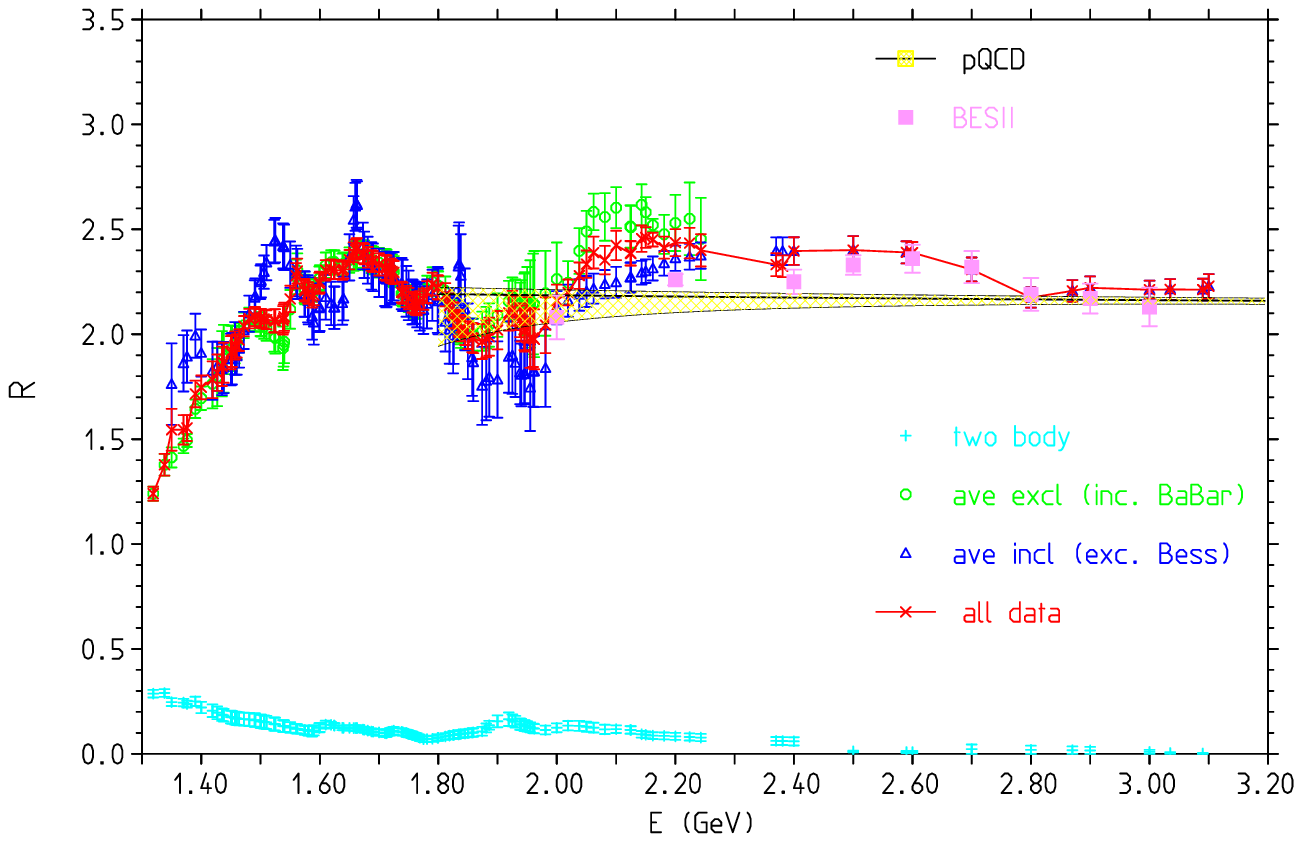}
\vspace*{-12mm}
\caption{Status of exclusive and inclusive measurements in the most
problematic region}
\label{fig:exvsinstatus} 
\vspace*{-6mm}
\end{figure}

At higher energies data are particularly problematic in the region
between 1.4 and 2.5 GeV. Fortunately, a new set of measurements is
available now from radiative return experiments at
BABAR~\cite{Aubert:2004kj} for the exclusive channels $e^+ e^- \to
\pi^+ \pi^- \pi^0,~ \pi^+ \pi^- \pi^+
\pi^-, ~ K^+ K^- \pi^+ \pi^-,~2(K^+ K^-),$ $~ 3\:(\pi^+ \pi^-),~
2(\pi^+ \pi^- \pi^0)~ \mathrm{and}~ K^+ K^-2(\pi^+ \pi^-)$.  These
data cover a much broader energy interval and extend to much higher
energies than previous experiments. The compilation of all data in
this range is shown in Fig.~\ref{fig:exvsinstatus}. Some of the early
experiments measured exclusive processes channel by channel, up to 2.5 GeV
including about 23 channels (see
Fig.\ref{fig:multi}) and others performed inclusive measurements for
$R(n>2)$ to which the two--body channels have to be added. The latter
drops to a small contribution above 1.4 GeV. This is in favor of
an inclusive strategy and helps to separate leptonic two prong events
from the hadronic channels dominated by $n>2$ events. In view of the many
channels an inclusive measurement seems to be more feasible at a
precision of about 1\% which should be attempted in the improvement
program. Table~\ref{tab:amualpchannels} gives a more detailed picture about
the relevance of the various modes for the contribution to the range
$2M_K < E < 2~\gv$.

In our analysis we are working throughout with renormalized bare cross
sections (\ref{Rundressed})~\cite{EJ95}.
For older data the missing renormalizations are performed
accordingly. Not performing these a posteriori radiative corrections
would lead to a value for $\amuh$ which is about 1 $\sigma$ higher.
Final state photon radiation for the low energy dominating $\pi^+\pi^-$ channel
is included as given by scalar QED. In fact this procedure is less
model dependent than it looks like in first place. The reason is that experiments
have subtracted radiative events using the same model, while in fact
at least the virtual hard photon effects are included in the measured
cross sections, as virtual effects cannot be eliminated by cuts~\cite{Hoefer:2001mx}.

\begin{table}[t]
\scriptsize
\centering
\caption{Contributions to $\amuh$ and $\dah0$ from the energy region $2M_K < E < 2~\gv$.
$X^*= X( \to \pi^0 \gamma)$, $_{iso}$=evaluated using iospin relations.}
\label{tab:amualpchannels}
\begin{tabular}{l|rr|rr}
\hline\noalign{\smallskip}
 channel $X$ & $\amu^X$ & \% & $\Delta \alpha^X$ & \% \\ 
\noalign{\smallskip}\hline\noalign{\smallskip}
 $  \pi^0\gamma                   $ &    0.04  &    0.04 &    0.00  &    0.03 \\
 $  \pi^+\pi^-                    $ &   11.99  &   11.66 &    1.59  &    9.64 \\
 $  \pi^+\pi^-\pi^0               $ &    9.22  &    8.98 &    1.25  &    7.55 \\
 $  \eta\gamma                    $ &    0.45  &    0.44 &    0.05  &    0.30 \\
 $  \pi^+\pi^-2\pi^0              $ &   19.27  &   18.75 &    3.79  &   22.93 \\
 $  2\pi^+2\pi^-                  $ &   13.99  &   13.62 &    2.80  &   16.92 \\
 $  \pi^+\pi^-3\pi^0 $~~$_{iso}$ &    1.17  &    1.14 &    0.26  &    1.56 \\
 $  2\pi^+2\pi^-\pi^0             $ &    1.94  &    1.88 &    0.43  &    2.60 \\
 $  \pi^+\pi^-4\pi^0 $~~$_{iso}$ &    0.08  &    0.08 &    0.02  &    0.12 \\
 $  \eta^*\pi^+\pi^-             $ &    0.26  &    0.25 &    0.05  &    0.31 \\
 $  2\pi^+2\pi^-2\pi^0            $ &    1.70  &    1.65 &    0.42  &    2.54 \\
 $  3\pi^+3\pi^-                  $ &    0.32  &    0.31 &    0.08  &    0.49 \\
 $  \omega^*\pi^0                $ &    0.77  &    0.75 &    0.13  &    0.78 \\
 $  K^+K^-                        $ &   21.99  &   21.39 &    2.64  &   15.94 \\
 $  K^0_SK^0_L                    $ &   13.17  &   12.82 &    1.49  &    8.99 \\
 $  \omega^*\pi^+\pi^-           $ &    0.09  &    0.08 &    0.02  &    0.12 \\
 $  K^+K^-\pi^0                   $ &    0.35  &    0.34 &    0.08  &    0.49 \\
 $  K^0_SK^0_L\pi^0 $~~$_{iso}$ &    0.35  &    0.34 &    0.08  &    0.49 \\
 $  K^0_SK^\pm\pi^\mp             $ &    1.08  &    1.05 &    0.25  &    1.49 \\
 $  K^0_LK^\pm\pi^\mp $~~$_{iso}$ &    1.08  &    1.05 &    0.25  &    1.49 \\
 $  K^+K^-\pi^+\pi^-              $ &    1.08  &    1.05 &    0.28  &    1.70 \\
 $  K\bar{K}\pi\pi $~~$_{iso}$ &    2.22  &    2.16 &    0.54  &    3.23 \\
  $  p\bar{p}                      $ &    0.07  &    0.07 &    0.02  &    0.12 \\
 $  n\bar{n}                      $ &    0.08  &    0.07 &    0.02  &    0.13 \\
   $  \phi \to {\rm \ missing}      $ &    0.03  &    0.03 &    0.00  &    0.02 \\
    sum                           &  102.78  &  100.00 &   16.54  &  100.00 \\
   {tot [sum in \%] }      &  692.00  &   [14.85] &   73.65  &   [22.46]  \\
\noalign{\smallskip}\hline
\end{tabular}
\end{table}

An intresting reconsideration of data in the $J/\psi$ resonance region
~\cite{Seth:2005ny} demonstrates very good agreement between
BES~\cite{BES} and much older Crystal Ball~\cite{Osterheld:1986hw} results and in
fact taking into account only the two data sets leads to a reduction
of the uncertainty in this region.

\begin{table*}[t]
\vspace*{-4mm}
\centering
\caption{Contributions for $\dahz \times 10^4$
(direct integration method) and
$\dah0 \times 10^4$ (non-perturbative part in the Adler function
method), with relative (rel) and absolute
(abs) error in percent.}
\label{tab:alphacont}
\begin{tabular}{crrr||rrr}
\hline\noalign{\smallskip}
   Energy range      & $\dahz \times 10^4$ &
  \tc{rel~[\%]}  & \tc{abs~[\%]} & $\dah0 \times 10^4$ &
  \tc{rel~[\%]} & \tc{abs~[\%]}  \\
\noalign{\smallskip}\hline\noalign{\smallskip}
   $\rho,\omega$ ($E<2M_K$)        &   36.23 [$\:    $ 13.1](0.24) &   0.7  &   1.1  &   33.29 [ 45.2](0.22) &   0.7 \% &   4.6 \% \\
        $2M_K<E<2~\gv$             &   21.80 [$\;\;\,$  7.9](1.33) &   6.1  &  34.9  &   16.44 [ 22.3](0.83) &   5.0 \% &  67.7 \% \\
     $2~\gv<E<M_{J/\psi}$          &   15.73 [$\;\;\,$  5.7](0.88) &   5.6  &  15.4  &    7.91 [ 10.7](0.44) &   5.6 \% &  19.3 \% \\
   $M_{J/\psi}<E<M_{\Upsilon}$     &   66.95 [$\:    $ 24.3](0.95) &   1.4  &  18.0  &   13.94 [ 18.9](0.29) &   2.0 \% &   8.1 \% \\
   $M_{\Upsilon}<E<E_{\rm cut}$    &   19.69 [$\;\;\,$  7.1](1.24) &   6.3  &  30.4  &    0.96 [  1.3](0.06) &   6.2 \% &   0.4 \% \\
   $E_{\rm cut}<E$ pQCD            &  115.66 [$\:    $ 41.9](0.11) &   0.1  &   0.3  &    1.09 [  1.5](0.00) &   0.1 \% &   0.0 \% \\
   $E < E_{\rm cut}$ data          &  160.41 [$\:    $ 58.1](2.24) &   1.4  &  99.7  &   72.55 [ 98.5](1.01) &   1.4 \% & 100.0 \% \\
           total                   &  276.07         [100.0](2.25) &   0.8  & 100.0  &   73.64 [100.0](1.01) &   1.4 \% & 100.0 \% \\     
\noalign{\smallskip}\hline
\end{tabular}
\end{table*}

Substantial progress is expected in this region by the future 
VEPP-2000~\cite{VEPP2000} at Novosibirsk and a possible new facility 
at Frascati~\cite{KLOE2}. Not only the effective fine structure constant
can be substantially improved, also the hadronic contribution to the
muon $g-2$ may be improved (see Tab.\ref{tab:amuhad})

\begin{table*}[t]
\vspace*{-1mm}
\centering
\caption{Contributions to
$a_\mu^{{\rm had}} \times 10^{10}$ with relative (rel) and absolute
(abs) error in percent.}
\label{tab:amuhad}
\begin{tabular}{crrr}
\hline\noalign{\smallskip}
               Energy range  &
  $a_\mu^{{\rm had}}$[\%]$(\rm error) \times 10^{10}$ & \tc{rel~[\%]} & \tc{abs~[\%]}  \\
\noalign{\smallskip}\hline\noalign{\smallskip}
   $\rho,\omega$ ($E<2M_K$)        &  538.33 [$\:$     77.8](3.65) &   0.7 \% &  42.0 \% \\
        $2M_K<E<2~\gv$             &  102.31 [$\:$     14.8](4.07) &   4.0 \% &  52.1 \% \\
     $2~\gv<E<M_{J/\psi}$          &   22.13 [$\;\;\,$  3.2](1.23) &   5.6 \% &   4.8 \% \\
   $M_{J/\psi}<E<M_{\Upsilon}$     &   26.40 [$\;\;\,$  3.8](0.59) &   2.2 \% &   1.1 \% \\
   $M_{\Upsilon}<E<E_{\rm cut}$    &    1.40 [$\;\;\,$  0.2](0.09) &   6.2 \% &   0.0 \% \\
   $E_{\rm cut}<E$ pQCD            &    1.53 [$\;\;\,$  0.2](0.00) &   0.1 \% &   0.0 \% \\
   $E < E_{\rm cut}$ data          &  690.57 [$\:$     99.8](5.64) &   0.8 \% & 100.0 \% \\
           total                   &  692.10 [        100.0](5.64) &   0.8 \% & 100.0 \% \\
\noalign{\smallskip}\hline
\end{tabular}
\end{table*}

\begin{figure}[t]
\vspace*{-0mm}
\centering
\includegraphics[height=10cm]{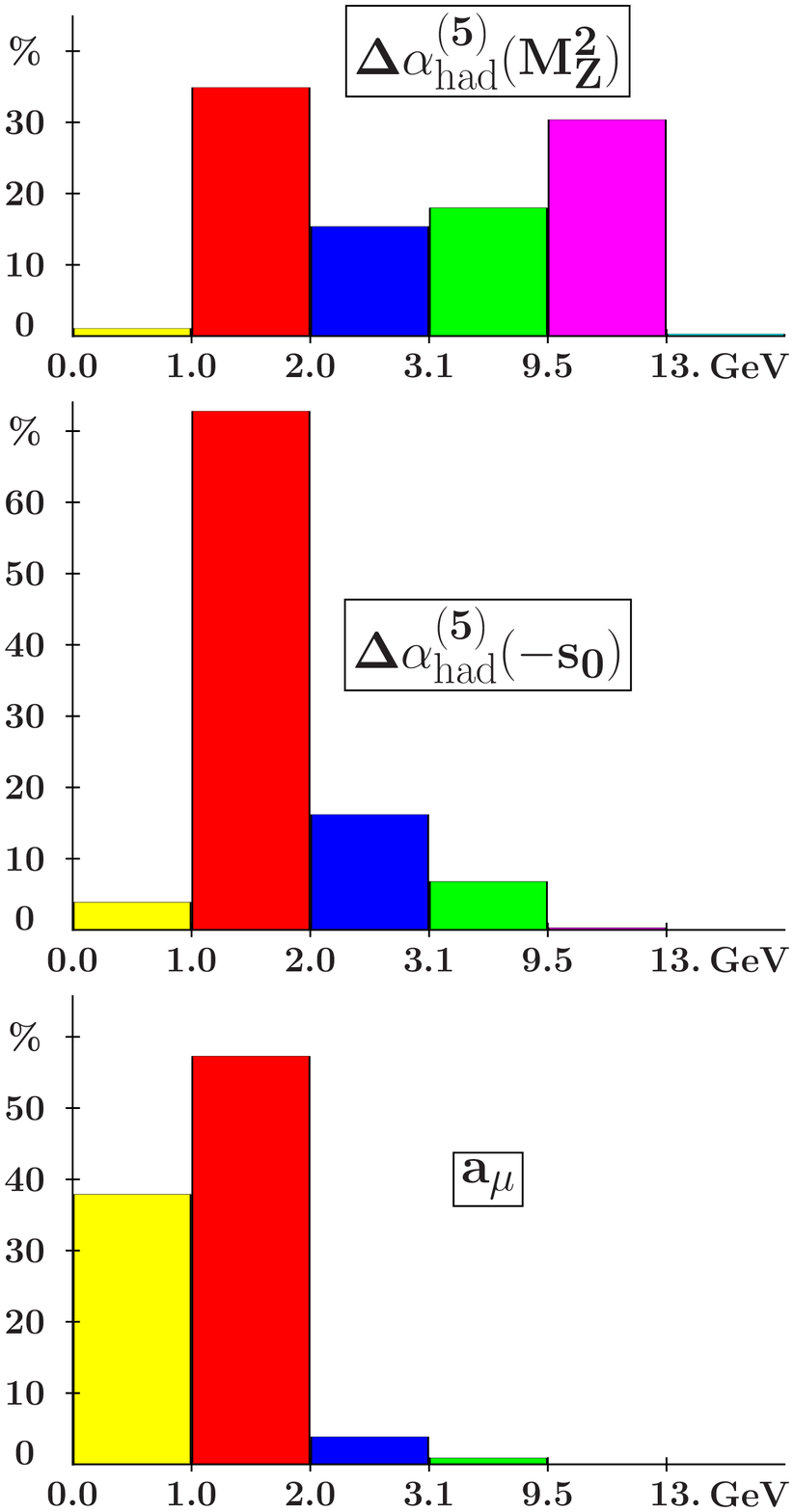}
\vspace*{-6mm}
\caption{Comparison of error profiles between $\dahz$, $\dah0$ and $a_\mu$ }
\label{fig:errdist} 
\vspace*{-6mm}
\end{figure}

The present error profile of the various quantities of interest is
compared in Fig.\ref{fig:errdist}.

\section{$\Delta \alpha^{\rm had}$ AND THE ADLER FUNCTION}
The Adler function is an ideal tool for 
disentangling perturbative from non--perturbative effects
in the Euclidean region. It is defined by
\bea
D_\gamma(-s) &\doteq&
\frac{3\pi}{\alpha}\:s\:\frac{d}{ds} \Delta \alpha(s) 
= -\left( 12 \pi^2 \right)\:s\: \frac{d\Pi'_{\gamma}(s)}{ds}
\eea
and thus for the hadronic part we may write
\ba
D_\gamma (Q^2)&=& Q^2\;\int\limits_{4 m_\pi^2}^{\infty}ds\,
\frac{R_\gamma (s)}{\left( s+Q^2 \right)^2}\epo
\ea
While the time--like function $R_\gamma(s)$ is calculable in pQCD
only by refering to quark--hadron duality, $D_\gamma(Q^2)$ is a smooth simple 
function both in terms of hadrons (dispersion integral over physical
cross sections) and in terms of quarks (pQCD) such that the validity
of pQCD can be examined directly. 
\begin{figure}[t]
\centering
\includegraphics[height=5cm]{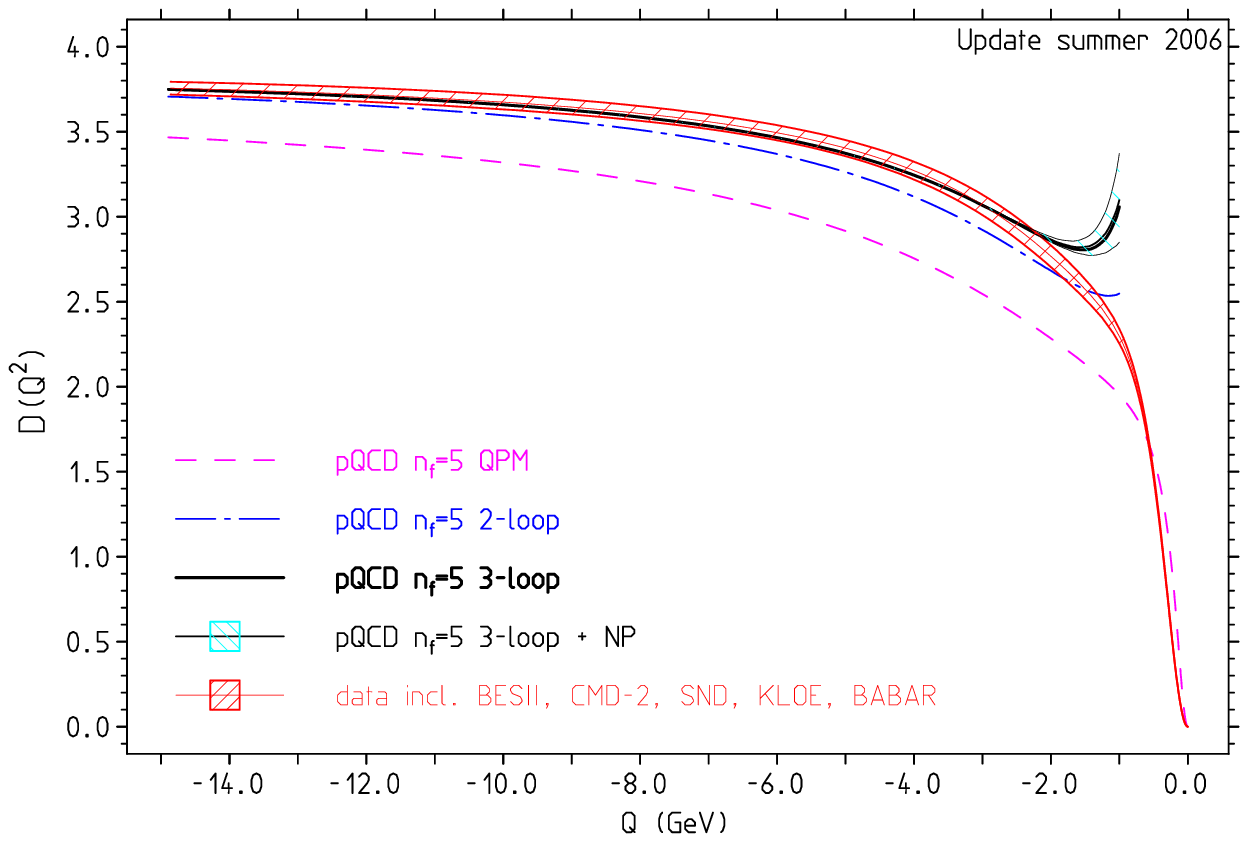}
\vspace*{-8mm}
\caption{``Experimental'' Adler-function versus theory (pQCD + NP)
in the low energy region (as discussed in~\cite{EJKV98}).
Note that the error includes both statistical and systematic ones, in
contrast to Fig.~\ref{fig:Rdata} where only statistcal errors are shown.}
\label{fig:Adler} 
\vspace*{-6mm}
\end{figure}
A comparison of the experimental
vs. the pQCD Adler function Fig.~\ref{fig:Adler} shows that pQCD in the Euclidean region
works very well for $\sqrt{Q^2}\gapprox 2.5~\gv$~\cite{EJKV98}. The
point is that at energies above that ``threshold'' massive QCD works
in any renormalization scheme provided mass effects are taken into
account correctly\footnote{Note that the curvature of $D_\gamma(Q^2)$ is
almost solely due to the quark masses. In massless QCD in the
\MSb scheme $D_\gamma(Q^2)$ is essentially a constant depending on the number
of flavors $N_f$.}. The breakdown of pQCD is obviously due the
fact that we are using the \MSb running coupling $\alsMSb (s)$ which grows
to $\infty$ at $\sqrt{s}=\Lambda_{\mathrm{QCD}}^{\overline{\rm
MS}}$\footnote{One could attempt to go to an infrared finite
renormalization scheme e.g. by redefining in pQCD
$$\alpha_s^{c-\mathrm{scheme}}(s)=\frac{\alsMSb (s)}{1+\alsMSb
(s)/c}\:,$$ where $c$ a constant $O(1)$. Now
$\alpha_s^{c-\mathrm{scheme}}(s)$ goes to the finite value $c$ (of
your chioce) as $\sqrt{s} \to \Lambda_{\mathrm{QCD}}^{\overline{\rm
MS}}$, which apparently would extend the validity of the pQCD result
to lower energies. But this would be at the expense of a huge scheme
dependence. The point is that $\alpha_s$ is not by itself an
observable and unless one precisely specifies an observable which
defines it, its meaning remains unclear. Also, this kind of
regularization ``by hand'' introduces an infrared fixed point in the
$\beta$-function which could be in conflict with confinement
expectations.  For a different point of view see~\cite{Milton:2005hp}
and references therein.}. In any case it looks convincing to calculate~\cite{FJ98}
\ba
\label{aleffsplitting}
\dahz&=&\dah0^\mathrm{data} \\
&&\hspace*{-1.5cm}+\left[\Delta\alpha^{(5)}_{\rm had}(-\MZ^2)-\dah0 \right]^\mathrm{pQCD} \crn
&&\hspace*{-1.5cm}+\left[\dahz-\Delta\alpha^{(5)}_{\rm
had}(-\MZ^2)\right]^\mathrm{pQCD} \nn
\ea
where the first term only has to be calculated from the experimental data.
The large perturbative second term now is rather sensitive to a
precise knowledge of the QCD parameters $\alpha_s$, $m_c$ and $m_b$, however.
For our pQCD evaluation we take the QCD parameters: \\
\centerline{$\alpha_s(M_Z)=0.1189\pm0.0010$~\cite{Bethke:2006ac},}
and quark masses (in GeV): \\[-3mm]
\scriptsize
\begin{center}
\begin{tabular}{cclcl}
\hline
\hline
masses & non-lattice &lattice  only   \\
\hline
$\bar{m}_c(\bar{m}_c)$ & $1.24\pm0.09$~\cite{Kuhn:2001dm} &
$1.30\pm0.03 \pm 0.20$~\cite{Rolf:2002gu}
\\
$\bar{m}_b(\bar{m}_b)$ & $4.20\pm0.07$~\cite{Kuhn:2001dm}
& $4.20\pm0.10\pm0.10$~\cite{Gimenez:2000cj} \\
\hline
\end{tabular}
\end{center}
\normalsize

\noi
as reviewed in~\cite{Yao:2006px}. Lattice results now provide
important cross checks of the phenomenological sum rule results which
we will actually use.

In this approach (see~\cite{FJLCnote01} for a detailed discussion of
the parameter dependences and error estimates) we obtain
\ba
\Delta\alpha^{(5)}_{\rm had}(-s_0)^{\mathrm{data}} &=& 0.007364 \pm
0.000101
\label{aleffs0}
\ea
and together with the second pQCD term we arrive at
\bea
\Delta\alpha^{(5)}_{\rm had}(-M_Z^2) &=& 0.027478 \pm 0.000144 \crn
\alpha^{-1}(-\mz) &=& 128.954 \pm 0.020\:.
\eea
Finally, for the third term of (\ref{aleffsplitting}) we obtain 
$$\Delta=0.000038\pm0.000002\;,$$ 
such that our final result reads
\ba
\Delta \al _{\rm had}^{(5)}(\mz) &=& 0.027547 \pm 0.000144 \crn
\alpha^{-1}(\mz) &=& 128.949 \pm 0.020\: .
\ea
We notice that this result already has a substantially lower error than the value obtained
via the direct dispersion integral. Errors of the perturbative part 
have been taken to be 100\% correlated (worst case). The sensitivity
to the charm mass in particular is substantial:\\[-4mm]
\scriptsize
\begin{center}
\begin{tabular}{cccc}
\hline
\hline
parameter & range & pQCD uncertainty  \\
\hline
$\alpha_s$ & 0.1179 ... 0.1199 & 0.000017  \\
$m_c$      & 1.15 ... 1.33 & 0.000078 \\
$m_b$      & 4.13 ... 4.27 & 0.000008 \\
\hline
\end{tabular}
\end{center}
\normalsize
but it can be reduced by choosing at higher scale like
$\sqrt{s_0}=10~\gv$ to diminish the $m_c$ dependence. The value
obtained is given in the abstract as my best conservative estimate. 

The conclusion of our analysis is that
while in the time-like approach pQCD works well only in the 
``perturbative windows''
$3.00 - 3.73$ GeV, $5.00 - 10.52$ GeV and $11.50 - \infty$,
in the space-like approach the plot of the Adler function 
shows that pQCD works well for all $Q^2=-q^2 > 2.5$ GeV.

For the future, in particular in view of ILC requirements, one should
attemt to improve the determination of the effective $\alpha_{\rm em}$
by a factor 10 in accuracy. What may we expect to be realistic? We assume that
dedicated cross section measurements are possible at the 1\% level
in the relevent energy regions. One then obtains the following estimates:  

\bmark with the direct integration of the data, and cuts applied as
above, 58\% of the hadronic contribution to $\alpha (M_Z)$ is
obtained from data and 42\% from pQCD. Given $\Delta \alpha_{{\rm
had}}^{(5)\:\mathrm{data}}
\times 10^4=160.41\pm 2.24$ (1.4\% error) assuming a 1\% overall
accuracy the error would be $\pm 1.60$. However, assuming that
different independent experiments are performed at
1\% accuracy for each region (divided up as in Tab.~\ref{tab:alphacont}) 
and adding up errors in
quadrature, one gets $\pm 0.83$. The improvement factor 
from the data  thus would be  2.7. The theory part 
{$\Delta \alpha_{{\rm had}}^{(5)\:\mathrm{pQCD}} 
\times 10^4=115.66\pm 0.11$} already now has an 0.1\% accuracy and
would not be required to be improved.

\bmark Using integration via the Adler function we have a 26\%
contribution from data and a 74\% pQCD. The experimental part is $\Delta \alpha_{{\rm had}}^{(5)\:\mathrm{data}} 
\times 10^4=73.64 \pm
1.01$ (1.5\% accuracy) and a 
1\% overall accuracy would reduce the error to $\pm 0.74$. Again,
assuming that independent 
1\% accuracy measurements are possible for each region (divided up as in Tab.~\ref{tab:alphacont}) 
and combining errors in
quadrature would yield $\pm 0.40$. Again the improvement factor is
essentially as above 2.8 but 5.6 with respect to the present direct
method which is usually adopted. The important point, however, is the very
different error profile Fig.~\ref{fig:errdist} in the two
approaches. In the Adler function approach errors may be reduced to a
large extent alone by low energy machines below 2.5 GeV.

\bmark One important draw back is that the large pQCD part 
$\Delta \alpha_{{\rm had}}^{(5)\:\mathrm{pQCD}} 
\times 10^4=201.83\pm 1.03$ uses pQCD down to much lower energies,
and the parameter uncertainties become much more severe leading to a 0.5\% accuracy
``only''. Here an improvement by a factor 5 would be desirable. There
has been steady progress in the past and we have no doubt that much
improvement will be possible in coming years.

\section{CONCLUSION}
The analysis presented above suggests that an improvement by a 
factor 5 in the error of \aleffZ 
may be realistic. The strategy for reaching this goal:

\bmark Adopt the Adler function approach to monitor and control
non-perturbative strong interaction effects, i.e., write $\dahz$
in the form (\ref{aleffsplitting}) and determine $\dah0$ from data
using the dispersion relation (\ref{aleffcut}).

\bmark The determination of $\dah0$ requires a dedicated program
of $\sha$ cross section measurements especially at low energy machines
attempting a precision of 1\% in the total cross section. As much as
possible the inclusive method should be pushed as it seems to be simpler
and easier to reach the 1\% accuracy. Detailed Monte Carlo simulation
and detector studies are necessary to clarify the possibilities.

\bmark Such experiments have to be accompanied by QED and SM calculations
of processes like Bhabha scattering, $\mu$--pair production and
$\pi$--pair production at least at the level of complete two--loop QED
plus resummations of leading higher order effects, and leading weak 
effects.

\bmark The evaluation of the missing pieces needed to obtain e.g. $\dahz$
may be performed using pQCD. This 74\% piece must be evaluated at a
precision at least at the 0.5\% level. The four--loop pQCD calculation
of the Adler function should be extended to include mass effects. Much
more important is a substantial improvement in the precise
determination of the QCD parameters $\alpha_s$, $m_c$, and, to a
lesser extent, $m_b$. Here lattice QCD must play a crucial role
(see e.g.~\cite{Sommer:2006wx} and references therein).

\bmark Obviously, to reach this goal requires a big effort especially on the
experimental side. Of course the simple straightforward direct
integration approach further will be applied and will provide an
important cross check for the evaluation based on the splitting
(\ref{aleffsplitting}). However, to reach the same precision using
this standard method would require much more experimental effort also
at higher energies as may be concluded from
Fig.~\ref{fig:errdist}.

In the near future progress will be possible by the radiative return
experiments at KLOE, BABAR and Belle~\cite{Shwartz:2005tp} up to
$\sqrt{s} \sim 3~\gv$ and by a new energy scan experiment at VEPP-2000
up to $\sqrt{s} = 2~\gv$. Further improvements of the accuracy of the
$R$ measurements in the range $3~\gv < \sqrt{s} < 5~\gv$ is also
expected from CLEO-C\cite{Dytman:2004mk} and from the $\tau$-charm
factory with BES-III~\cite{Harris:2006qh} in Beijing. At Frascati
a new facility DANAE/KLOE-2 planned could start data taking in 2010
and provide a further indispensable step in the improvement program on
\aleffZ  and $\amuh$. Note that a 1\% measurement in the range 1 to 2 GeV
would reduce the error of (\ref{aleffs0}) to
$\Delta\alpha^{(5)}_{\rm had}(-s_0)^{\mathrm{data}} = 0.007364 \pm
0.000060$.

A challenging long term project for lattice QCD is the direct
determination of the Adler function from first principles. First
attempts were made in the past~\cite{Gockeler:2000kj}, however, the
non-perturbative part is largely dominated by the proper inclusion of
the $\pi\pi \to \rho $ resonance. The dominant low energy part is extremely
sensitive to mass and width of the $\rho$~\cite{GJ04}, and to get
them correct requires to perform simulations at the
physical parameters of coupling and light quark masses.\\[0mm]

\noi
{\bf Acknowledgments}\\ 

Special thanks go to A. Denig, S. Eidelman,
W. Kluge, K. M\"onig, S. M\"uller, F. Nguyen, G. Pancheri and
G. Venanzoni for numerous stimulating discussions. I am very grateful
to H. Meyer, O. Tarasov and G. Venanzoni for carefully reading the manuscript.  I
also thank Frascati National Laboratory and the KLOE group for the
kind hospitality extended to me. This work was supported in part by
DFG Sonderforschungsbereich Transregio 9-03 and by the European
Community's Human Potential Program under contract HPRN-CT-2002-00311
EURIDICE and the TARI Program under contract RII3-CT-2004-506078.

\end{document}